\documentclass[useAMS,usenatbib]{mn2e}
\usepackage{graphicx}
\usepackage{amssymb}
\usepackage{amsmath}
\usepackage{algpseudocode}
\usepackage{multicol}
\usepackage{longtable}
\usepackage{algorithm}
\usepackage{cprotect}

\title[Primordial star formation]{Impact of an accurate modeling of primordial chemistry in high resolution studies}

\author[S. Bovino et al.]{ S. Bovino\thanks{Corresponding author: sbovino@astro.physik.uni-goettingen.de}$^1$, T. Grassi$^2$, M. A. Latif$^1$, and D. R. G. Schleicher$^1$\\
$^1$Institut f\"ur Astrophysik Georg-August-Universit\"at, Friedrich-Hund Platz 1, 37077 G\"ottingen\\
$^2$Department of Chemistry, Sapienza University of Rome, P.le A. Moro, 00185, Rome}
\begin{document}
\newcommand{\ith}{$i$th }
\newcommand{\jth}{$j$th }
\newcommand{\nth}{$n$th }

\newcommand{\dd}{\mathrm d}
\newcommand{\mA}{\mathrm A}
\newcommand{\mB}{\mathrm B}
\newcommand{\mC}{\mathrm C}
\newcommand{\mD}{\mathrm D}
\newcommand{\mE}{\mathrm E}
\newcommand{\mH}{\mathrm H}
\newcommand{\mSi}{\mathrm Si}
\newcommand{\mO}{\mathrm O}
\newcommand{\cmc}{\mathrm{cm}^{-3}}
\newcommand{\real}{\mathbb R}
\newcommand{\superscript}[1]{\ensuremath{^{\scriptscriptstyle\textrm{#1}\,}}}
\newcommand{\trader}{\superscript{\textregistered}}

\newcommand\mnras{MNRAS}
\newcommand\apj{ApJ}
\newcommand\aap{A\&A}
\newcommand\apss{Ap\&SS}

\date{Accepted *****. Received *****; in original form ******}

\pagerange{\pageref{firstpage}--\pageref{lastpage}} \pubyear{2013}

\maketitle

\label{firstpage}

\begin{abstract}
The formation of the first stars in the Universe is regulated by a sensitive interplay of chemistry and cooling with the dynamics of a self-gravitating system. As the outcome of the collapse and the final stellar masses depend sensitively on the thermal evolution, it is necessary to accurately model the thermal evolution in high resolution simulations. As previous investigations raised doubts regarding the convergence of the temperature at high resolution, we investigate the role of the numerical method employed to model the chemistry and the thermodynamics. Here we compare the standard implementation in the adaptive-mesh refinement code \verb|ENZO|, employing a first order backward differentiation formula (BDF), with the 5th order accurate BDF solver \verb|DLSODES|.  While the standard implementation in \verb|ENZO| shows a strong dependence on the employed resolution, the results obtained with \verb|DLSODES| are considerably more robust, both with respect to the chemistry and thermodynamics, but also for dynamical quantities such as density, total energy or the accretion rate. We conclude that an accurate modeling of the chemistry and thermodynamics is central for primordial star formation.
\end{abstract}

\begin{keywords}
cosmology: early Universe -- methods: numerical.
\end{keywords}

\section{Introduction}

According to hierarchical paradigm of structure formation, the first stars are formed as a result of gravitational collapse in minihalos of $\rm \sim10^5-10^6$~M$_\odot$ at  $z\sim20-30$ \citep{Abel2002, Bromm04, Yoshida08}. The formation and masses of these stars are relevant due to their radiative, chemical and mechanical feedback  on the environment \citep{Ciardi05, Tornatore07, Schleicher08, Schneider08}. While the first simulations pointed towards masses of more than $100$~M$_\odot$, more recent studies suggest efficient fragmentation in self-gravitating protostellar disks, implying that the stellar masses might be substantially reduced \citep{ClarkGlover2011, Greif2011ApJ, Smith11, Greif2012}. Simulations including radiative feedback indeed suggest upper mass limits of the order $\sim 60$~M$_\odot$ \citep{Hosokawa11,Susa2013}.

As fragmentation in turbulent gas clouds depends sensitively on the thermodynamics of the gas \citep{LiKlessenMacLow03,Peters2012}, an accurate modeling of chemistry and cooling is thus necessary. In a high-resolution study pursued with the adaptive mesh refinement code \verb|ENZO|, \citet{Turk2012} reported a strong dependence of the thermodynamics and chemistry on the resolution per Jeans length, which resulted in strong variations of physical quantities such as the radial velocities and the accretion rates. While local quantities like the detailed morphologies may indeed depend on the resolution in a turbulent self-gravitating system \citep{Sur10, Federrath11, Latif13}, radially averaged properties are typically expected to have only a minor resolution dependence. At the same time, \citet{Greif2013a} reported only a weak resolution dependence in calculations of primordial star formation, employing the moving-mesh code \verb|AREPO| \citep{Springel2010} and a high-order scheme for solving the chemistry and thermodynamics. Understanding the origin of this discrepancy is crucial, and in this letter, we will demonstrate that this behavior is closely related to the solver employed for solving the chemical network.


One of the most commonly used chemistry solvers in the context of primordial star formation is the first order BDF scheme  proposed by \citet{Anninos97}, which involves two central approximations aimed at reducing the computational time: (i) some of the species (e.g. H$^{-}$ and H$_2^+$) are assumed to instantaneously reach the equilibrium due to their short reaction timescales and are decoupled from the non-linear system (ii) the thermal evolution, i.e., the cooling and the heating functions are evaluated outside the BDF scheme and are used to determine the maximum chemical time-step instead of being solved non linearly within the BDF scheme. In the release of ENZO v2.2, we note that the second approximation was relaxed significantly by iterating between the chemistry and cooling module on a time-step that is the minimum of the chemical and cooling timestep.

In \citet{TurkScience2009,Turk2010ApJ} an improvement of the simple 1st order BDF method has been presented. A \verb|TWOSTEP| 2nd-order BDF solver based on a Gauss-Seidel technique including an Aitken extrapolation to accelerate convergence has been employed \citep{Verwer1994}. All the rate equations have been solved without applying equilibrium criteria. 

More sophisticated solvers have also been used in private and public versions of the \verb|FLASH| code \citep{Micic2012MNRAS,GloverMicic2013,Gray2010ApJ,Gray2013}. The latter includes two higher-order solvers compared to the standard BDF method:  the semi-implicit multi-order Bader-Deuflhard solver \citep{Bader1983} and the 3rd order Rosenbrock method \citep{Rentrop1979} that while being more performant is not very well suited for system with dimensions $N>10$ \citep{NumericalRecipes}.  
 
The hydrodynamics code \verb|ZEUS-MP| \citep{Glover2007ApJS,Glover2007ApJ,Glover2010}, and cosmological codes \verb|GADGET| \citep{Jappsen2007ApJ,ClarkGlover2011} and \verb|AREPO| \citep{Springel2010,Greif2011ApJ,Greif2012} employ the \verb|VODE| class solvers \citep{Brown1989}, both in their FORTRAN (\verb|DVODE|) and C (\verb|CVODE|) versions, which belong to the well-established \verb|ODEPACK/SUNDIALS| package\footnote{computation.llnl.gov/casc/sundials/main.html} \citep{Hindmarsh2005} that allows to solve stiff ODE systems more efficiently. 

While \verb|DVODE| fits astrophysical problems as well, it can be inefficient when the system is represented by a sparse Jacobian. In that case, the \verb|DLSODES| solver \citep{Hindmarsh83}, which has the capability of handling sparse matrices, is more efficient. In astrophysical applications,   chemical networks frequently consist of a sparse or a very sparse Jacobian matrix (a Jacobian primarily populated with zeros). The performances of  \verb|DLSODES| has been already discussed in \citet{Nejad2005}, and the efficiency of the two solvers has been compared for larger network \citep{Wakelam2008} in previous studies \citep{Grassi2011,Grassi2013}. 

In this letter, we will demonstrate the importance of choosing accurate chemistry solvers like \verb|DLSODES|, and show how the latter leads to converged radially averaged profiles for physical quantities like the radial velocity or the accretion rate. The structure of this letter is as follows. In Section \ref{sec:models}, we describe the initial setup and our new implementation of the chemistry in the version 2.2 of cosmological code  \verb|ENZO|. Results for various resolutions of 16, 32, 64 and 128 cells per Jeans length are presented in section \ref{sec:results}. We compare our main results with earlier studies \citep{Turk2012} and present our conclusions in section \ref{sect:conclusions}.

\begin{figure}
\begin{center}
	\includegraphics[width=.4\textwidth]{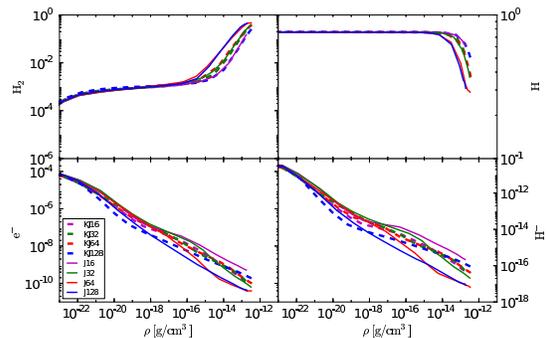}
	\caption{Chemical species mass fraction for different resolutions. Both original enzo (marked as J) and our new implementation (marked as KJ) results are shown. From the left top to right bottom panels: molecular hydrogen, atomic hydrogen, electron and H$^-$ fractions.}\label{fig:1}
\end{center}
\end{figure}

\begin{figure*}
\begin{center}
	\includegraphics[width=.8\textwidth]{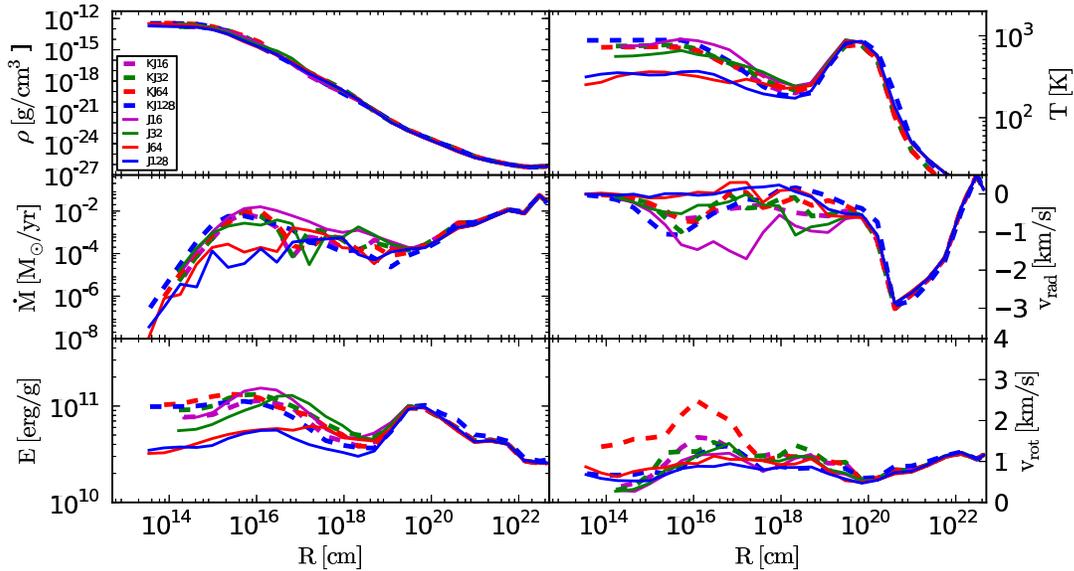}
	\caption{Spherically averaged radial profiles of the 8 simulations taken at roughly the same peak density. Upper left is the the total density in g cm$^{-3}$, upper right is the temperature in K, middle left the accretion rate in M$_{\odot}/yr$, middle right is the radial velocity in km s$^{-1}$, bottom left the total energy density in erg/g and bottom right the rotational velocity in km s$^{-1}$. Note that radial and rotational velocities have been evaluated subtracting the bulk velocity (for details see \citep{LatifBH}).}\label{fig:2}
\end{center}
\end{figure*}

\section{Numerical methods}\label{sec:models}

The simulations are performed with the publicly available version 2.2 of the \verb|ENZO| code. They are started from cosmological initial conditions generated from Gaussian random fields.  Our simulations start at redshift z = 99 with a top grid resolution of 128$^3$ cells. Two initial nested levels of refinement are subsequently added each with a grid resolution of 128$^3$ cells. Our simulation box has a cosmological size of 0.3 Mpc h$^{-1}$ and is centered on the most massive minihalo. In total, we initialize 5767168 particles to compute the evolution of the dark matter dynamics and have a final dark matter resolution of 70 $\rm M_{\odot}$. The parameters for creating the initial conditions and the distribution of baryonic and dark matter components are taken from the WMAP seven years data \citep{Jarosik2011}. We further allow additional 27 levels of refinement in the central 18 kpc region of the halo during the course of simulation. It gives us a total effective resolution of 0.9 AU in comoving units. The resolution criteria used in these simulations are based on the Jeans length, the gas over-density and the particle mass resolution. We stop the simulations once they reach the maximum refinement level. A split hydro solver with a 3rd order piece-wise parabolic (PPM) method for hydrodynamical calculations is employed. The dark matter N-body dynamics is solved using the particle-mesh technique, and a multigrid Poisson solver is employed for the self-gravity computations.

\begin{figure*}
\begin{center}
	\includegraphics[width=.7\textwidth]{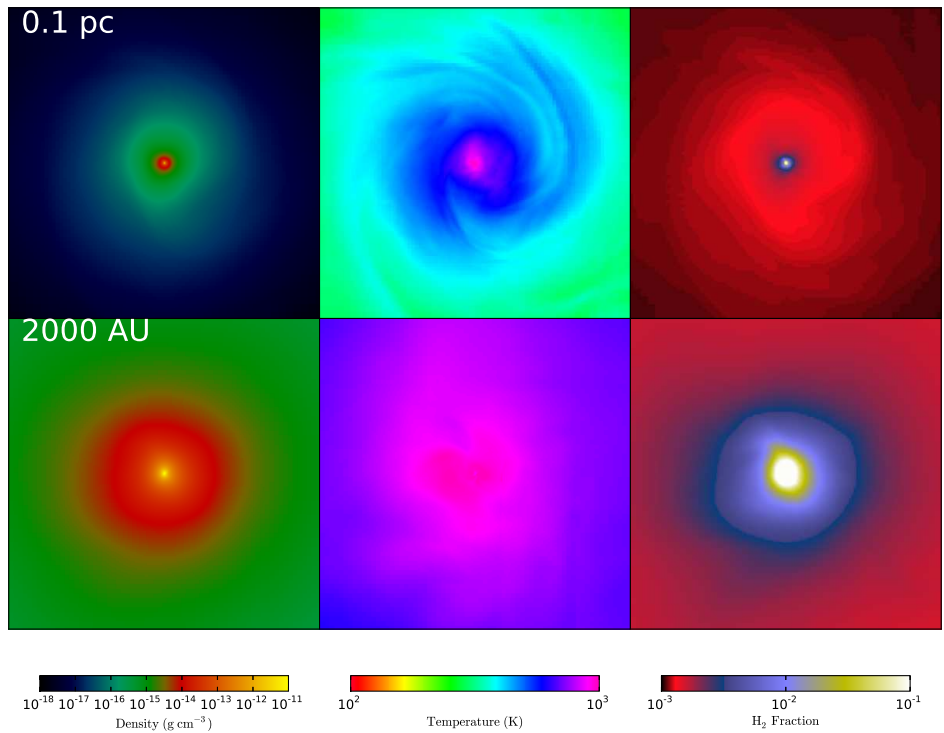}
	\caption{The slices for the density (left columns),  temperature (second columns) and molecular hydrogen mass fraction (right columns) for KJ128 run are shown at fields of view of 0.1, and 2000 AU from the top to bottom.}\label{fig:4}
\end{center}
\end{figure*}

\subsection{Chemical package}\label{sec:our}
In our new implementation we adopt the scheme included in the new chemical package \verb|KROME|, which will be presented in a forthcoming paper. It consists of a new chemical framework which includes accurate solvers, heating and cooling functions,  rate equations routines and  reduction techniques for chemical networks (see \citet{Grassi2011,Grassi2013})  that allows its applicability in a number of different astrophysical environments. All  physical and computational ingredients to solve for the chemical and thermodynamical evolution come from the package and are included in  \verb|ENZO| through a call to the \verb|KROME| routine.

The temperature is evolved alongside the rate equations and  is updated at each internal solver's time-step leading to a more accurate solution. The solver we decided to employ in the package is the \verb|DLSODES| aimed at dealing with sparse Jacobian networks as the one employed in the present calculations (sparsity$\sim$62\%).

\subsection{Chemistry}\label{sec:chemistry}

We follow the non-equilibrium evolution of 9 species: H, H$^+$, H$^-$, H$_2$, H$_2^+$, He, He$^+$, He$^{2+}$, and e$^-$. In total, 20 kinetic reactions have been included same as available with Enzo code \citep{Turk2012}. The formation of H$_2$ is very sensitive to the choice of 3-body reaction as pointed out by \citet{Turk2011ApJ}. Here we adopt the three-body rates of \citet{Abel2002} for all of our runs, both in the standard \verb|ENZO| and in our \verb|KROME| implementation. 

The system of coupled ODEs that takes into account the reactions which form and destroy the \ith species is self-consistently solved within cosmological simulations. Heating and cooling include H$_2$ formation heating/cooling due to the 3B reactions (4.48 eV/molecule) as described in \citet{Omukai2000}, H$_2$ cooling as reported by \citet{GloverAbel08}, bremsstrahlung, H and He collisional ionization, excitation and recombination cooling as given by \citet{Cen92}. The temperature change due to the heating and cooling is calculated as
\begin{equation}
	\frac{dT}{dt} = \frac{\gamma -1}{k_B\sum_in_i}(\Gamma - \Lambda) \hspace{0.2cm} \mathrm{K/s}.
\end{equation}
We note here that this equation is integrated simultaneously with the chemical rate equation both in \verb|KROME| and \verb|ENZO| 2.2.
\section{Results}\label{sec:results}

In all, we have carried out 8 simulations to study the thermal evolution of minihaloes which are potential sites for the formation of the first stars.  We compare our new implementation with the one employed in  \verb|ENZO|~2.2. The Jeans resolutions of 16, 32, 64 and 128 cells (hereafter called J16, J32, J64, J128 for standard runs and KJ16, KJ32, KJ64, KJ128  for  \verb|KROME|) were mandated throughout the evolution of the simulations.  The abundances of various chemical species for different Jeans resolutions are shown in Fig.~\ref{fig:1} both for \verb|KROME| and  \verb|ENZO|~2.2.  On a qualitative level, the results from both solvers approximately follow our expectations. In particular, the electron density shows a characteristic power-law decrease towards high densities, while the H$_2$ abundance grows significantly at densities of $\sim 100$~cm$^{-3}$ and again at densities of $\sim10^{10}$~cm$^{-3}$ due to the three-body reactions. However, the results obtained with the  \verb|ENZO|~2.2 implementation show a significant amount of variation, while the \verb|KROME| results are close to each other and show a considerably improved convergent behavior. 
We note that the J64  and J128 runs in the \verb|ENZO|~2.2 implementation were stopped earlier due to non-convergence in the chemical solver while our results reached higher densities. For the sake of clarity, the final results are plotted for the same peak density allowing a more consistent comparison between the different runs. 

The dynamical quantities for the same peak density are shown in Fig.~\ref{fig:2}. The density profiles follow an almost isothermal behavior and are similar for various runs. Larger differences are present in the  \verb|ENZO|~2.2 runs, in particular at densities above $\rm 10^6$~cm$^{-3}$ due to the discrepancies in the H$_2$ and electron abundances. The \verb|KROME| results, on the other hand, are almost identical. Consequently, the total energy and accretion rates are similar for different resolutions but relatively large variations are found for the  \verb|ENZO|~2.2 runs.  Overall, better convergence is found for the radial velocity in \verb|KROME|, even though we note a small deviation in the J128 run, where the radial velocity is slightly higher. The rotational velocities are about 1 km/s for all runs at radii larger than 0.1 pc. The variations in the center may arise due to the differences in the morphologies, as we always expect some fluctuations in turbulent self-gravitating systems.  

Fig.~\ref{fig:4} shows the density distribution, temperature and the H$_2$ abundance for our highest resolution run (i.e., KJ128). It is noted that the H$_2$ abundance largely follows the density, and also the temperature reflects the underlying density field. Due to the large density range shown here, the density distribution shows only minor deviations from spherical symmetry, as the global collapse dynamics are dominating the evolution. Clear deviations from spherical symmetry and signs of rotation can however be recognized in the temperature, as the overall contrast between high-density and low-density regions is now decreased. Similar features can be seen in the H$_2$ abundance, which is however increasing towards the central core.

We focused here on \verb|ENZO|~2.2 for illustrative purposes, but note that we performed additional tests with \verb|ENZO| version 2.0, 2.1, 2.2 and 3.0 (still in development). Their overall behavior is quite similar, implying a temperature drop above some critical resolution. \verb|ENZO|~3.0 appears to be slightly more stable up to  a  resolution of  64 cells per Jeans length, but shows a similar behavior at higher resolution.

\section{Discussion and conclusions}\label{sect:conclusions}
Overall, our results suggest significant improvements in the convergence behavior when the high-order \verb|DLSODES| solver implemented in the \verb|KROME| package is employed for the \verb|ENZO| simulations. We then relate the non-convergent behavior reported by \citet{Turk2012} to the first order BDF method. In addition, our results agree with the findings by \citet{Greif2013a}, who also employed a higher-order accurate method for the chemistry. We thus conclude that an accurate modeling of the chemistry is crucial for high-resolution studies of primordial star formation.

We expect the latter to be particularly relevant in situations of strong cooling, i.e. where the temperature changes significantly with density. This effect can be further enhanced in the presence of metals or dust \citep{Omukai2005, Schneider06, Dopcke13} and also in situations where H$_2$ cooling is initially suppressed by photo-dissociating backgrounds, but becomes relevant at higher densities \citep[e.g.]{Shang09, Schleicher10b, Latif12}.


As previous studies pointed towards a need of high resolutions per Jeans length for an accurate modeling of turbulence and turbulent fragmentation \citep{Sur10, Federrath11, Turk2012, Latif13, LatifBH}, we here highlight the importance of using high-order BDF methods such as \verb|DLSODES| for modeling the chemical evolution. The package \verb|KROME| will be published in a companion paper, including a general framework to construct chemical networks, specific examples, code interfaces and will be released for the public use. 


\section*{Acknowledgements}
S.B. and D.R.G.S. thank for funding through the DFG priority program `The Physics of the Interstellar Medium' (projects SCHL 1964/1-1). D.R.G.S. and M.L. thank for funding via the SFB 963/1 on "Astrophysical Flow Instabilities and Turbulence". T.G. gratefully acknowledges the CINECA consortium for the financial support. The simulation results are analyzed using
the visualization toolkit for astrophysical data YT \citep{Turk2011a}. We are grateful to the anonymous referee for the useful comments.

\bibliographystyle{mn2e}      
\bibliography{mybib_new}


\bsp

\label{lastpage}

\end{document}